\documentclass[11pt]{article}
\newcommand{\spc}{\hspace{0.22in}}

\usepackage{graphicx}
\def \ds {\displaystyle}

\def \ns {\normalsize}

\def \es {\enspace}
\def \ts {\thinspace}

\title{
  \vspace*{-1.5cm}
  \hfill{\ns KEK-TH-877} \\ 
  \vspace*{-0.3cm}
  \hfill{\ns April 2003}  \\
  \vspace*{2.0cm}
  {\Large {\bf Theory of Membrane in Heegaard Diagram Expansion}} \\
  \vspace*{0.8cm}
}
\author{ Hirotaka Sugawara\thanks{e-mail: sugawara@phys.hawaii.edu, 
hirotaka.sugawara@kek.jp} }
\date{}
\begin{document}
\maketitle
\begin{center}
%\vspace{5mm}
%{\bf KEK, High Energy Accelerator Research Organization}
%{\bf Tsukuba, Ibaraki, Japan} Address after April 1, 
{\it Department of Physics and Astronomy, University 
of Hawaii at Manoa, Hawaii, U.S.A.}
\end{center}

\vspace{3cm}
%
%=== Abstract
%
\begin{abstract}
 The vacuum amplitude of the closed membrane theory is investigated using the 
fact that any three-dimensional manifold has the corresponding Heegaard diagram
(splitting) although it is not unique. We concentrate on the topological aspect
with the geometry considered only perturbatively. In the simplest case where 
the action describes the free fields we find that the genus one amplitudes 
(lens space) are obtained from the $S^3$ amplitude by merely renormalizing the
membrane tension. The amplitudes corresponding to the Heegaard diagram of genus
two or higher can be calculated as the Coulomb amplitudes with arbitrary charge
distributed on a knot or a link which corresponds to the set of branch points 
of a given regular or an irregular covering space.  
We also discuss the case of membrane instanton.  
\end{abstract}
\clearpage
%
%================================================================
\section{Introduction}
%================================================================
\spc Although the M theory seems to be occupying one corner of the moduli space
of all the string theories~\cite{Polchinski}, it still lacks a formulation 
which is as intuitively comprehensible as the other string theories. TypeIIB 
theory is known to have both the string and the matrix formulations~\cite{IKKT}
and the M theory has so far only the matrix formulation~\cite{BFFS}. The latter
is known to have some connection to the membrane theory~\cite{WHN} but its 
legitimacy as the fundamental theory is in doubt~\cite{HW}.

  Whether the theory is fundamental or just effective it may happen that the 
vacuum amplitude is dominated by the process of all possible closed membranes 
being created and annihilated. If it is true, then, the scattering amplitude 
of closed membranes can be calculated just by inserting some vertex operators 
before the path integration is performed for the vacuum amplitude.  

  To locate a vertex operator in some finite position in the Euclidean space, 
the conformal invariance was essential in the case of string theory. We 
implicitly assume that the same should be true also in the case of membrane 
theory, although the essential part of our discussions does not depend on the 
form of the action. In the case of closed string (fig.~1~(a)), the vacuum 
amplitude is obtained by summing all the amplitudes corresponding to 
fig.~1~(b), namely all the amplitudes for the closed two-dimensional surfaces.
In the case of closed membrane theory we must consider the processes which 
involve all types of closed surfaces shown in fig.~1~(b). These surfaces should
be created and annihilated in the vacuum process. The surfaces with a higher 
genus may be regarded as a bound state of lower genus surfaces but instead we 
treat them on an equal footing in the approach described here. The process of 
creation and the subsequent propagation, or the splitting etc.~and finally the 
annihilation of a closed string is described by an amplitude on a closed 
surface. This should be extended also to the case of closed surface. This means
that the vacuum amplitude of the closed membrane theory must be the sum of 
amplitudes of all possible three-dimensional closed manifolds.  

  It is well known that the torii with genus number $g=0, 1, 2, \ldots, \infty$
exhaust all possible two-dimensional connected orientable surfaces. The 
three-dimensional case is more complicated. Here we rely on the fact that any 
closed, orientable three-dimensional manifold has at least one Heegaard diagram
(or Heegaard splitting): let $H_1$ and $H_2$ be solid torii of the same genus 
$g$,and let $h:\partial H_2 \to \partial H_1$ be a homeomorphism. Then any 
closed connected orientable three-dimensional manifold $M^3$ can be identified 
with $\displaystyle H_1 \bigcup_h H_2$;
\begin{center}
$M^3=\displaystyle H_1 \bigcup_h H_2$.
\end{center}
There is no uniqueness of the diagram as is shown in fig.2 where $S^3$ is shown
to have a Heegaard diagram ($H_1,\ts H_2,\ts h$) with $H_1$ and $H_2$ of any 
number 
of genus $g$. This causes some interesting complications which can be solved 
if the Poincar\'e conjecture is valid for the three-dimensional manifolds as 
will be explained in the subsequent sections.  

  The application of membrane theory to physics may not come from the 
fundamental theory (even if it exists) but rather from the solitonic membrane.
The process of membranes appearing and disappearing in the Euclidean time is 
nothing but the membrane instanton~\cite{BBS}.  In some models, this is the only
source of four-dimensional superpotential and, thus, this may be the reason for
the smallness of masses of quarks and leptons~\cite{Witten}.  The issue will be 
briefly discussed in the subsequent sections although, admittedly, a lot will 
remain in the future investigation. 

  Section 2 will be devoted to the general discussions and mathematical 
preliminaries. Section 3 will deal with the case of $g=1$ which corresponds to 
the lens spaces. Section 4 treats the case of $g=2$ and higher. Section 5 
contains summary and discussions.  
%
%================================================================
\section{General discussions}
%================================================================
\spc Much of our discussions are independent of the action we adopt for the 
membrane.  Still, the concrete results seem to depend on it rather crucially 
because of the geometrical nature of our results.  Let us start from the 
following expression for the generic vacuum amplitude:  
\begin{equation}
F= \sum_{{\mit\Sigma}_i} \int e^{-s} \displaystyle \prod_\alpha 
d \psi_\alpha \es,
\label{eq:vcmAmp}
\end{equation}
where ${\mit\Sigma}_i$ stands for all possible closed connected orientable
three manifolds, and the action $S$ is of the form:
\begin{equation}
S= \int_{{\mit\Sigma}_i} L_m d^3 \sigma \es,
\label{eq:actnS}
\end{equation}
with an appropriate membrane Lagrangian $L_m$. $\psi_\alpha$ stands for the 
generic field which comes into the Lagrangian $L_m$. We can adopt as $L_m$ the 
super symmetric Lagrangian given in ref.~\cite{Witten} or, for simplicity, 
the following Weyl invariant bosonic membrane Lagrangian as an effective 
Lagrangian, 
\begin{equation}
L_m =\frac{\ds T_m}{\ds 2} 
%\int d^3 \sigma 
\sqrt{\det g_{\alpha \beta}} \ts 
\phi \ts \left( g^{\alpha \beta} \partial _\alpha X^\mu \partial_\beta X_\mu 
- \phi^2 \right) \es,
\label{eq:LgrgnMmbrn}
\end{equation}
where $T_m$ is the membrane tension, $g_{\alpha \beta} \es (\alpha , \beta = 
1,2,3)$ is the metric tensor, $X^\mu (\sigma) \es (\mu = 1, \ldots , D)$ is the
membrane coordinate in the $D$-dimensional target space and $\phi$ is the 
compensating field. Both the target space and the world volume space are taken 
to be Euclidean.  Whether the Lagrangian (\ref{eq:LgrgnMmbrn}) or its 
supersymmetric extension can be free of Weyl anomaly for certain value of $D$ 
is itself an interesting problem but we will not touch on this issue here. Our 
main concern in this work is the sum over ${\mit\Sigma}_i$. 

  As is explained in the introduction each ${\mit\Sigma}_i$ has at least one 
Heegaard diagram $(H_1,\ts H_2,\ts h)$.  The fig.~3~(a) illustrates the case of
$g=1$ and the fig.~3~(b) corresponds to the case of $g=2$. Each case can be 
transformed to a more physically intuitive fig.~4~(a) and fig.~4~(b),
respectively. In fig.~4~(a), a thin torus surface is created at the Euclidean 
time $t=t_1$ and propagates within $H_1$ until $t=0$.  At $t=0$, all sorts of 
$h$ transform the torus surface into $H_2$. The surface propagates within $H_2$
until it disappears at $t=t_3$.  Similar interpretation can be given to 
fig.~4~(b) with $g=2$ torus. The $g=0$ case is omitted here as the simplest 
exercise for the reader.  

  The whole problem, therefore, comes down to the issue of classification of 
$h$ in each genus case. This, of course, is basically solved in the well 
established theorem due to Lickorish~\cite{Lickorish} which states that any $h$ 
for genus $g$ Heegaard diagram is isotopic to the composition of the Dehn 
twists along $3g-1$ closed curves. The curves for $g=1$ and $g=2$ cases are 
already given in fig.~3~(a) and fig.~3~(b) as $C_1, C_2$ and $C_1, \ldots , 
C_5$ respectively and the general case in fig.~5. Although much is known for 
this problem~\cite{CB}, it still looks formidable to perform the summation in
eq.~(\ref{eq:vcmAmp}) for all possible $h$. Fortunately, there seem to be ways 
out as will be explained in the following sections for each value of $g$. The 
method is different in each case of $g$ so that the discussion will be left to 
the sections which deal with individual cases.  

  In each case we cannot perform the integration for $g_{\alpha \beta}$. 
We concentrate on the topological aspect and we pick a particular value for 
$g_{\alpha \beta}$ and try to perturb $g_{\alpha \beta}$ around this value.  
Instanton case is slightly different in that it is independent to 
$g_{\alpha \beta}$ from the beginning and we must consider effectively the 
following case~\cite{BBS},
\begin{equation}
L_m =  \int_{\mit\Sigma} C
\label{eq:crtnThrForm}
\end{equation}
where $C$ is a certain three form defined on ${\mit\Sigma}$ which is a 
three-dimensional manifold in a certain seven-dimensional compact space in 
case of M theory. $C$ can be written as 
\begin{equation}
C = \displaystyle \sum_{i =1}^{b^{(3)}} a_i \omega_i
\end{equation}
where $\omega_i$ is a component of a basis of three forms and $b^{(3)}$ is
the third Betti number of ${\mit\Sigma}$.  These can be calculated in principle
for a given Heegaard diagram $(H_1,\ts H_2,\ts h)$. We touch on this case in 
the following separate sections for $g=1$ and for $g \ge 2$.  

  In the process of adding the contributions from $g=0, 1, 2, \ldots$ we 
encounter the problem of double counting due to the non-uniqueness of the 
Heegaard diagram although it is meaningless until we know exactly how to 
determine the coefficient in front of each contribution. In principle, this 
can be solved assuming the validity of the Poincar\'e conjecture which states 
that the homology group together with the fundamental group classifies the 
topology of the manifolds. Fox example, $g=0$ is $S^3$ and we know out of $g=1$
Heegaard diagrams there exists only one with the trivial fundamental group.  
We, therefore, subtract this contribution from the $g=1$ contribution.  This 
kind of renormalization or subtraction procedure can in principle be applied 
to higher genus cases although Poincar\'e conjecture in three dimensional case 
is yet to be proven beyond $g \ge 3$.  
%
%===========================================================================
\section{$g=1$ case}
%===========================================================================
\spc The Heegaard diagram $(H_1,\ts H_2,\ts h)$ for $g=1$ is well understood. 
Each $h$ corresponds to the following $SL_2 (Z)$ operation 
\begin{equation}
\left(
 \begin{array}{c}
   {i_2}\\
   {j_2}
 \end{array}
\right)
=
\left(
 \begin{array}{cc}
   {q^\prime} & p \\
   {p^\prime} & q 
 \end{array}
\right)
\left(
 \begin{array}{c}
   {i_1}\\
   {j_1}
 \end{array}
\right), \es
q q^\prime - p p^\prime = 1 \es.
\label{eq:SL2Z}
\end{equation}
Here $(i_1 , j_1)$ or $(i_2 , j_2)$ corresponds to the 
$i_1 \ts (i_2)$-th meridian or to the $j_1 \ts (j_2)$-th 
longitude respectively. The reducible case is 
$q + q^\prime = \pm 2$.\footnote{We are not certain whether we should include 
irreducible ones in our summation.}

  The products of Dehn twists correspond to $(i_1, j_1) = ( 1, 0)$ 
in eq.~(\ref{eq:SL2Z}):
\begin{equation}
\left(
 \begin{array}{c}
   {i_2}\\
   {j_2}
 \end{array}
\right)
=
\left(
 \begin{array}{cc}
   {q} & {p^\prime} \\
   {p} & {q^\prime} 
 \end{array}
\right)
\left(
 \begin{array}{c}
   1 \\
   0
 \end{array}
\right) 
=
\left(
 \begin{array}{c}
   q \\
   p
 \end{array}
\right) \es.
\label{eq:DhnTwst}
\end{equation}
Since $(H_1,\ts H_2,\ts h^{-1}) = (H_1,\ts H_2,\ts h)$, we should calculate what sort 
of product of Dehn twists corresponds to $h^{-1}$.  We easily see that 
\begin{equation}
\left(
 \begin{array}{cc}
   q & p^\prime\\
   p & q^\prime
 \end{array}
\right)^{-1}
\left(
 \begin{array}{c}
   1\\
   0 
 \end{array}
\right)
=
\left(
 \begin{array}{c}
   q ^\prime \\
   -p
 \end{array}
\right) 
\simeq
\left(
 \begin{array}{c}
   q^\prime \\
   p
 \end{array}
\right) \es,
\end{equation}
where $\simeq$ means the equivalence as a Heegaard diagram or the topological 
equivalence. Together with eq.~(\ref{eq:SL2Z}) this implies
\begin{equation}
L(p,\ts q) = L(p, \ts q^\prime) \es {\rm when} \es qq^\prime = 1 \es
({\rm mod} \ts p) \es, 
\label{eq:HgrdDhn}
\end{equation}
where $L(p, q)$ signifies the Heegaard diagram corresponding to the product of 
Dehn twists given in in eq.~(\ref{eq:DhnTwst}). Brody's theorem~\cite{Brody} 
states that the converse is true: The only topologically equivalent case for 
$L(p,\ts q)$ is given by eq.~(\ref{eq:HgrdDhn}) except for the following 
trivial cases, 
\begin{equation}
L(p,\ts q) \cong L(p,\ts -q) \cong L(-p,\ts q) \cong L(-p,\ts -q) \cong 
L(p,\ts q + k p) \es.
\label{eq:DhnTrvl}
\end{equation}
$L(p, q)$ is known to be a lens space.  The correspondence between the lens 
space and the Dehn twist is illustrated in fig.~6~(a) $\sim$ fig.~6~(d) where 
we treat the case of $L(5,\ts 2)$~\cite{Rolfsen}.  

  We now work with the unit ball depicted in fig.~6~(a) where a surface point 
on the northern hemisphere after an eastward rotation of an angle equal to 
$2 \pi q / p$ is identified with the southern point with the same latitude.  
The $q = 1$ case corresponds to $S^3 / Z_p$ and was used by 
E.~Witten~\cite{Witten} to construct a model for the doublet-triplet splitting 
without the twist the sector problem which accompanies the orbifold 
cases~\cite{KHMN}. The fundamental group of $L(p,\ts q)$ is $Z_p$ independently 
of $q$ and, therefore, can be used instead of $S^3 / Z_p$ for the purpose of 
doublet-triplet splitting although the physical implication is not clear.  

  As an example, let us now try to compute $F$ given in eq.~(\ref{eq:vcmAmp})
for the case of ${\mit\Sigma}_i = L(p,\ts q)$ with the action given in 
eq.~(\ref{eq:LgrgnMmbrn}). We adopt the $\phi=1$ gauge and assume that the 
$g_{\alpha \beta}$ can be perturbed around the metric corresponding to 
$L(p,\ts q)$ of fig.~6~(a).  
$F$ can be written in the following form:
\begin{equation}
F = \displaystyle \sum_{p, q} \int e^{-\frac{D-2}{2} T_m \int d^3 \sigma 
\partial_\alpha X \partial_\alpha X} \displaystyle \prod_\sigma 
d x(\sigma) \es,
\label{eq:exprnF}
\end{equation}
where the boundary condition for $X(\sigma) = X (r,\ts \theta,\ts \varphi)$ 
with the usual spherical coordinate $(r,\ts \theta,\ts \varphi)$ is given by 
\begin{equation}
X (1,\ts \theta,\ts \varphi) = X(1,\ts \pi - \theta,\ts 
\varphi + \frac{2 \pi q}{p}) \es. 
\label{eq:bndrCndtn}
\end{equation}
It is easy to see that the expression given in eq.~(\ref{eq:exprnF}) is 
independent of $q$. Assuming that we do not need any $q$-dependent arbitrary 
constant before summing over $q$, we have 
\begin{equation}
F = \sum_p n_p e^{- \frac{D-2}{2} T_m \int d^3 \sigma \partial_\alpha 
X \partial_\alpha X} \prod_\sigma d x(\sigma) \es,
\label{eq:FbfrSum}
\end{equation}
where $n_p$ is the number of $q\es (< p)$ with $q$ and $p$ having no common 
divisor. The boundary condition (\ref{eq:bndrCndtn}) turns into a restriction 
on the coefficients in the expansion 
\begin{equation}
X(r,\ts \theta,\ts \varphi) = \sum_{n=1}^\infty \sum_{m=-n}^n 
a_{n, m} (r) Y^m _n (\theta,\ts \varphi) \es, 
\end{equation}
where $Y^m _n (\theta, \varphi)$ is the usual spherical harmonics.  
We find 
\begin{equation}
n + \left( 1+ \frac{\ds 2q}{\ds p} \right) m = 2N \es,
\end{equation}
where $N$ is an arbitrary integer.  
This leads to the following restrictions: 
\begin{eqnarray*}
& & \underline{\mbox{For odd } p} \hspace*{15cm} \\
& & \mbox{if } kp \le n \textless (k+1)p \mbox{ then}, \\
& & \underline{\mbox{for even n}},\es m=0,\ts \pm 2p,\ts \ldots,\ts 
\pm (k) p, \\
& & \underline{\mbox{for odd n}},\es m = \pm p,\ts \pm3p,\ts \ldots,\ts 
\pm [k]p \es,
\end{eqnarray*}
where $(k)$ is the largest even number smaller than or equal to $k$ and $[k]$ 
is the largest odd number smaller than or equal to $k$. Similar restriction can
be obtained for even $p$. The integration in (\ref{eq:FbfrSum}) can be done in 
an elementary way and we obtain,
\begin{equation}
F = \displaystyle \sum_{p,q} F \left( L(p,\ts q),\ts T_m \right) =
\sum_p n_p F (S^3 / Z_p,\ts T_m) = \sum_p n_p F(S^3,\ts T^p _m) \es, 
\label{eq:cntrbtnMfd}
\end{equation}
where $F(A,\ts T_m)$ is the contribution to $F$ in eq.~(\ref{eq:FbfrSum}) from 
a given manifold $A$ with the membrane tension $T_m$.  We have the expression
\begin{equation}
T^p _m = T_m \frac{\ds \sum\nolimits_k \Bigl( k(k+1)+1 \Bigr)}
{\ds \sum\nolimits_k \Bigl( kp(kp+1)+1 \Bigr)} 
\cong \ts p^{-2} T_m \es. 
\label{eq:tnsnMmbrn}
\end{equation}
We may have a $p$-dependent constant $\alpha_p$ multiplied to each contribution
in equation (16). The number $n_p$, therefore, may not mean anything but we 
list first several numbers in Table~\ref{tab:frstSvrl_N_p} anyway. 
\begin{table}[htbp]
\vspace{0.6cm}
\begin{center}
\begin{tabular}{|l|l|l|l|l|l|l|l|l|l|l|l|l|l|l|l|}
\hline
p & 1 &2 &3 &4 &5 &6 &7 &8 &9 &10 &11 &12 &13 & 14 & $\ldots$ \\
\hline
$n_p$ &1 &1 &2 &2 &3 &2 &5 &4 &5 &5 &9 &4 &11 &6 & $\ldots$ \\
\hline
\end{tabular}
\end{center}
\vspace{-0.5cm}
\caption{Values of first several $n_p$}
\label{tab:frstSvrl_N_p}
\vspace{0.3cm}
\end{table}

  Equation~(\ref{eq:cntrbtnMfd}) together with (\ref{eq:tnsnMmbrn}) shows that 
the contribution from $L(p,\ts q)$ can be calculated from that of $S^3$ simply 
by renormalizing the membrane tension. Since $p=1$ corresponds to $S^3$ either 
we subtract this from eq.~(\ref{eq:cntrbtnMfd}) or we omit the contribution 
from $g=0$ which also corresponds to $S^3$. These discussions are relevant only
when we know how to compute $\alpha_p$ on the basis of the unitarity or some
other physical principles. We may take a limit of thin membrane and relate this
to the string theory, where unitarity determines $\alpha_p$. The case of 
membrane instanton (eq.~(\ref{eq:crtnThrForm})) is much easier for 
$L(p,\ts q)$. We take the transverse coordinate to be vanishing in 
eq.~(\ref{eq:crtnThrForm}). Then we have 
\begin{equation}
L_m = \int_{L(p, q)} P d^3 \sigma
\end{equation}
where $P$ is a pseudoscalar density.  This leads to 
\begin{equation}
F(L(p,\ts q)) = F(S^3) \es,
\end{equation}
for all values of $p$ and $q$.
%
%================================================================
\section{$g \ge 2$ case}
%================================================================
\spc We start with the case of $g=2$ Heegaard diagram $(H_1,\ts H_2,\ts h)$ 
where $h$ corresponds to a product of any number of matrices each factor 
corresponding to one of five Dehn twists along the curves $C_1,\ldots, C_5$ in 
fig.~3~(b).  Each Dehn twist can be expressed as a $4 \times 4$ matrix since 
only four out of five curves $C_1,\ldots, C_5$ constitute a basis. What is 
remarkable in the $g=2$ case is that we can choose $C_1,\ldots, C_5$ and, 
therefore, all the Dehn twists to be invariant under the $180^\circ$ 
rotation around the$X$ axis shown in fig.~3~(b).

  This fact leads to a theorem by Birman and Hilden~\cite{Birman} which states 
that a $g=2$ Heegaard diagram $(H_1,\ts H_2,\ts h)$ is topologically equivalent
to a 2-fold branched coverings of $S^3$ with a knot or a link as the branch 
set.  The proof is illustrated in fig.~7~(a) to 
fig.~7~(d)~\cite{Honma}.

  The generic expression for this case can be written as, 
\begin{equation}
F= \int e^{-s_1 -s_2} \prod_{\tilde{\sigma} \in L} \delta \bigl( \psi^{(1)} 
(\tilde{\sigma}) - \psi^{(2)} (\tilde{\sigma}) \bigr) \prod_\sigma d 
\psi^{(1)} (\sigma)d \psi^{(2)} (\sigma) \es, 
\end{equation}
where $s_i = \int_{S^3 _i} L(\psi^{(i)} (\sigma)) d \sigma$ and $L$ 
stands for a given link. $\psi^{(i)}(\sigma)$ is a field which comes into the 
Lagrangian corresponding to the $i$-th sphere $S^3_i$.  

  First, let us consider the action in eq.~(\ref{eq:LgrgnMmbrn}). 
We take $\phi = 1$ gauge and treat $g_{\alpha\beta}$ perturbatively as in the 
case of $g=1$. We have, 
\begin{equation}
F= \int \exp\left[ - \frac{\ds D-2}{\ds 2} T_m \sum^2 _{i=1} \int_
{S^3_i} d^3 \sigma \partial_\alpha X^{(i)} \partial_\alpha X^{(i)} \right] 
\cdot \prod_{\tilde{\sigma} \in L} \delta \bigl( X^{(1)} (\tilde{\sigma})
- X^{(2)} (\tilde{\sigma}) \bigr) \prod_\sigma dX^{(i)} (\sigma) \es.  
\end{equation}
The integration can be done in an elementary way with the result, 
\begin{equation}
F= \int \exp\left[(D-2) T_m \int d^3 \sigma E_\alpha (\vec{\sigma}) E_\alpha 
(\vec{\sigma})\right] \prod_s dk(s) \es,
\label{eq:rsltF}
\end{equation}
where $\partial_\alpha E_\alpha = {\ds \int_{\vec{\sigma}} (s) \in L} ds \ts 
\delta (\vec{\sigma} (s) - \vec{\sigma}) K (s)$. $s$ is a parameter along the 
link $L$. This means that we are calculating nothing but a Coulomb amplitude 
with all possible charge distribution on the link. Eq.~(\ref{eq:rsltF})
can be rewritten as, 
\begin{eqnarray}
F &=& \int \exp\left[ -\frac{(D-2)T_m}{4 \pi} \int_{\vec{\sigma}(s) \in L} ds 
\int_{\vec{\sigma}(s^\prime) \in L} ds^\prime 
\frac{k(s)k(s^\prime)}{\mid \vec{\sigma}(s) - \vec{\sigma} (s^\prime) \mid} 
\right] \prod_s dk(s) \nonumber \\
  &=& \sum_L \left[ \det \left( \frac{1}{\mid \vec{\sigma} (s) 
- \vec{\sigma} (s^\prime) \mid} \right) \right]^{-1/2} \ts,
\label{eq:sumL}
\end{eqnarray}
where $\vec{\sigma}(s), \vec{\sigma} (s^\prime) \in L$. The summation is over 
all possible links. We obtained the result for the flat metric. 

  It will be very interesting to see what we will get for the generic metric. 
Our approach is to use perturbation theory for $g_{\alpha \beta}$.
To avoid the double counting we need to subtract from eq.~(\ref{eq:sumL})
a contribution from manifold which has the same homology and the fundamental 
group corresponding to $L(p,\ts q)$ treated in the previous section. We should
be able to obtain these for each $L$ in principle. 

  Let us now proceed to discuss the case of $g \ge 3$. Here we depend on the 
Alexander-Hilden-Montesinos theorem~\cite{AHM}, the Hilden and Montesinos 
version of which states that every closed orientable 3-manifold $M$ is an 
irregular 3-fold branched covering of $S^3$ branched over a knot. The 
irregularity means the lack of symmetry which we have in the case of $g=2$.  
This does not matter for our purposes. Moreover, since the theorem applies to 
any three-manifold, the calculation applies to any value of $g$.  

  The result obtained is similar to the case of $g=2$ except that we have two 
kinds of charges distributed on the knots, 
\begin{equation}
F= \int \exp\left[ -\frac{(D-2) T_m}{4\pi} \int_{\sigma \in L} 
\frac{k_1 (s) k_1(s^\prime)-k_1 (s) k_2 (s^\prime) + k_2 (s) k_2 (s^\prime)}
{\mid \vec{\sigma}(s) - \vec{\sigma}(s^\prime) \mid} ds ds^\prime \right] 
\prod_s dk(s) \es.
\label{eq:arbgnsFlt}
\end{equation}
Although this formula can be applied to any value of $g$, this does not 
necessarily undermine the formulae for $g=1$ and $g=2$ which have been 
performed previously since they are much simpler. Again we note that 
eq.~\ref{eq:arbgnsFlt} was obtained using the flat metric.  

  The membrane instanton for the case $g \ge 2$ does not seem to allow us to 
proceed without the knowledge of explicit form for basis three forms 
$\omega^{(3)} _{i j k, \alpha}$ in the following formula, 
\begin{eqnarray}
F= \int d k (s) \exp\biggl[ i \int d^3 \sigma \hspace*{-4mm}
& &\hspace*{-5mm} \biggl\{ p^{(1)} (\sigma) \int ds \delta (\vec{\sigma} - 
\vec{\sigma}(s)) p^{(1)} (\sigma) k (s) \nonumber \\
&+&\hspace*{-2mm} p^{(2)} (\sigma) + \int ds \delta (\vec{\sigma} 
- \vec{\sigma}(s)) p^{(2)} (\sigma) k (s) \biggr\} \biggr] \es,
\end{eqnarray}
where $p= \varepsilon^{i j k} (\displaystyle \sum^{b_3 (X)} _{\alpha =1} 
a_\alpha \omega_{i j k, \alpha})$. This is the formula for $g=2$ and we have a 
similar formula for $g \ge 2$. These formulae are quite useless unless we know 
what $\omega$'s are ($a_\alpha$ can be integrated as a four dimensional field).
%
%================================================================
\section{summary and discussions}
%================================================================
\spc The vacuum amplitude for the membrane theory has been discussed. We 
started 
with the non-unique classification of three-dimensional manifolds using the 
Heegaard diagram. This gives some intuitive understanding of what is going on
in the vacuum amplitude. That is, a thin torus of genus $g$ is created and 
becomes fatter. At certain stage it gets twisted in a homeomorphic manner and 
propagates until it disappears. A Heegaard diagram $(H_1,\ts H_2,\ts h)$ for 
$g=1$ with reducible $h$ is a lens space $L(p,\ts q)$. It has been shown that 
the vacuum amplitude contribution is the same as $S^3$ with the membrane 
tension renormalized. We have restricted all our calculations to zeroth order 
in $h_{\alpha \beta}$ where $g_{\alpha \beta} = g^0 _{\alpha \beta} + 
h_{\alpha \beta}$.  The contribution from $g \ge 2$ is described as a Coulomb 
amplitude with one ($g=2$) or two ($g \ge 3$) kinds of arbitrary charges 
distributed on the knot or the link which plays the role of branch set of 
covering spaces. We have the two-fold covering space of $S^3$ for the case of 
$g=2$ and three-fold covering space of $S^3$ for $g \ge 3$. 
Eq.~(\ref{eq:sumL}) indicates that our result is a geometric one.  
Fig.~8~(a) and fig.~8~(b) contribute more or less the same amount to $F$. 
Both give a determinant of a matrix in eq.~(\ref{eq:sumL}) with large diagonal 
elements and large matrix elements where $s$ and $s^\prime$ are both close to 
either of $A$, $B$ and $C$.  This indicates that not only the local but also 
global nature of the knot may be important.  The summation over 
$g_{\alpha \beta}$ should play an important role in understanding the 
situation. Eq.~(\ref{eq:arbgnsFlt}) is valid for any three dimensional manifold
with no reference to Heegaard diagram. This means in particular that we can 
use eq.~(\ref{eq:arbgnsFlt}) for the cases of $g=1$ or $g=2$. But still the 
Heegaard diagram seems to provide us a means to understand the situation in an 
intuitive manner.  It may also happen that smaller genus contributions provide 
us a good approximation.  What we have done in this article is quite 
preliminary and more investigation has to be done especially to 
consider the generic $g_{\alpha \beta}$.  
\vspace*{0.8cm}

\noindent{\large \bf Acknowledgment} \\
I would like to thank H.~Hagura for checking the
manuscript.
\vspace*{1cm}

\begin{figure}[htbp]
\begin{center}
\includegraphics[width=16cm]{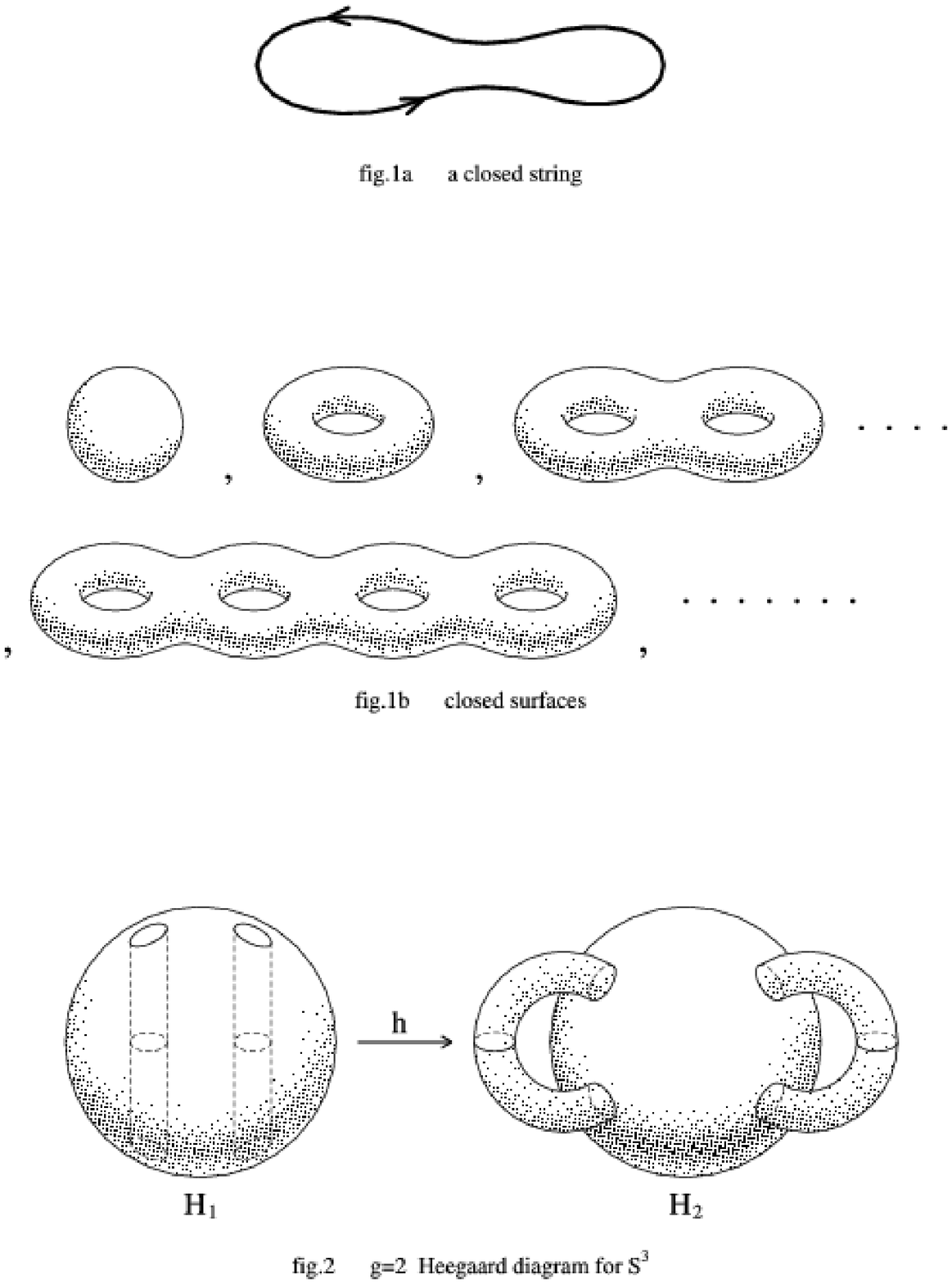}
\end{center}
\end{figure}

\begin{figure}[htbp]
\begin{center}
\includegraphics[width=16cm]{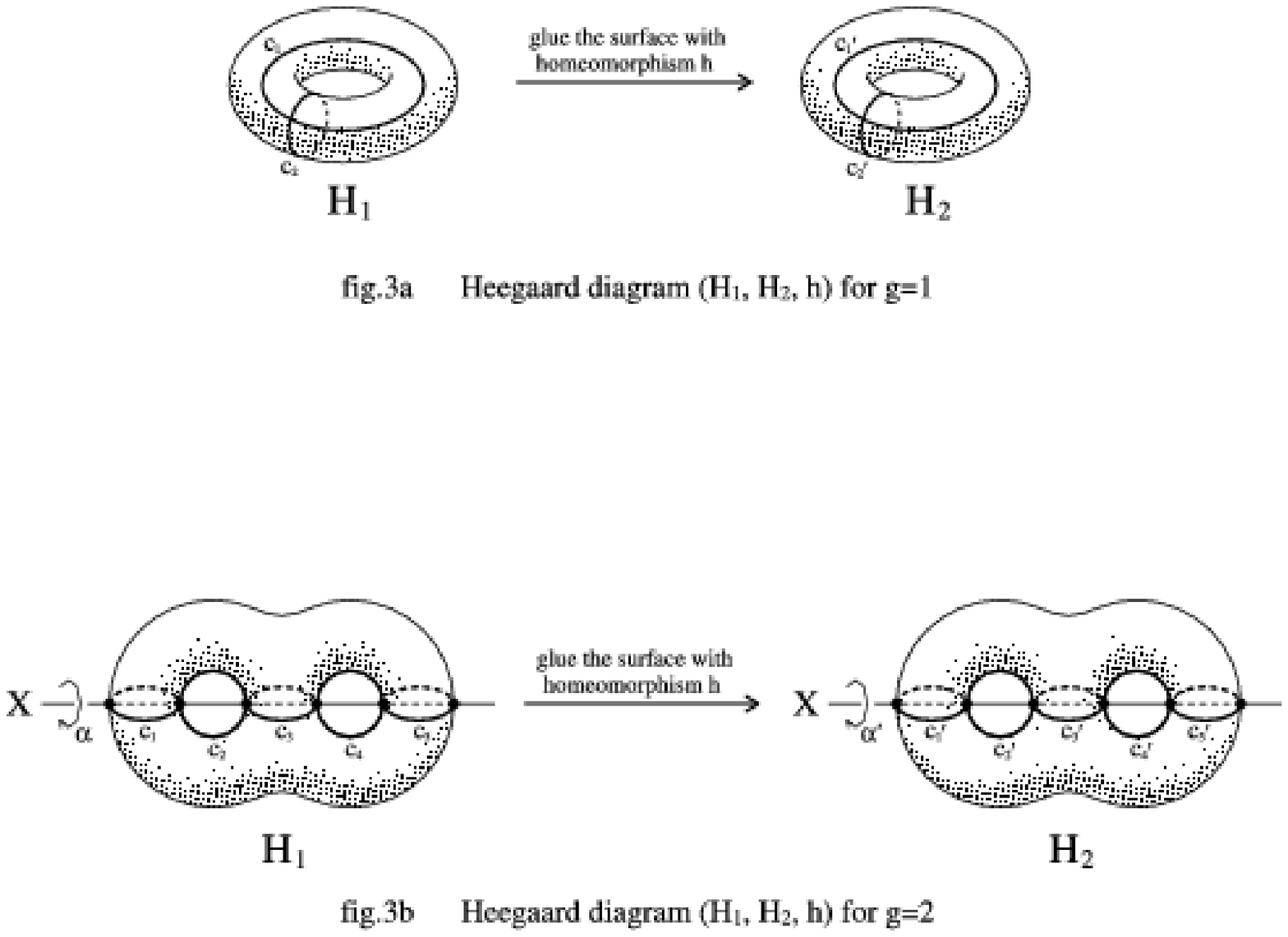}
\end{center}
\end{figure}

\begin{figure}[htbp]
\begin{center}
\includegraphics[width=16cm]{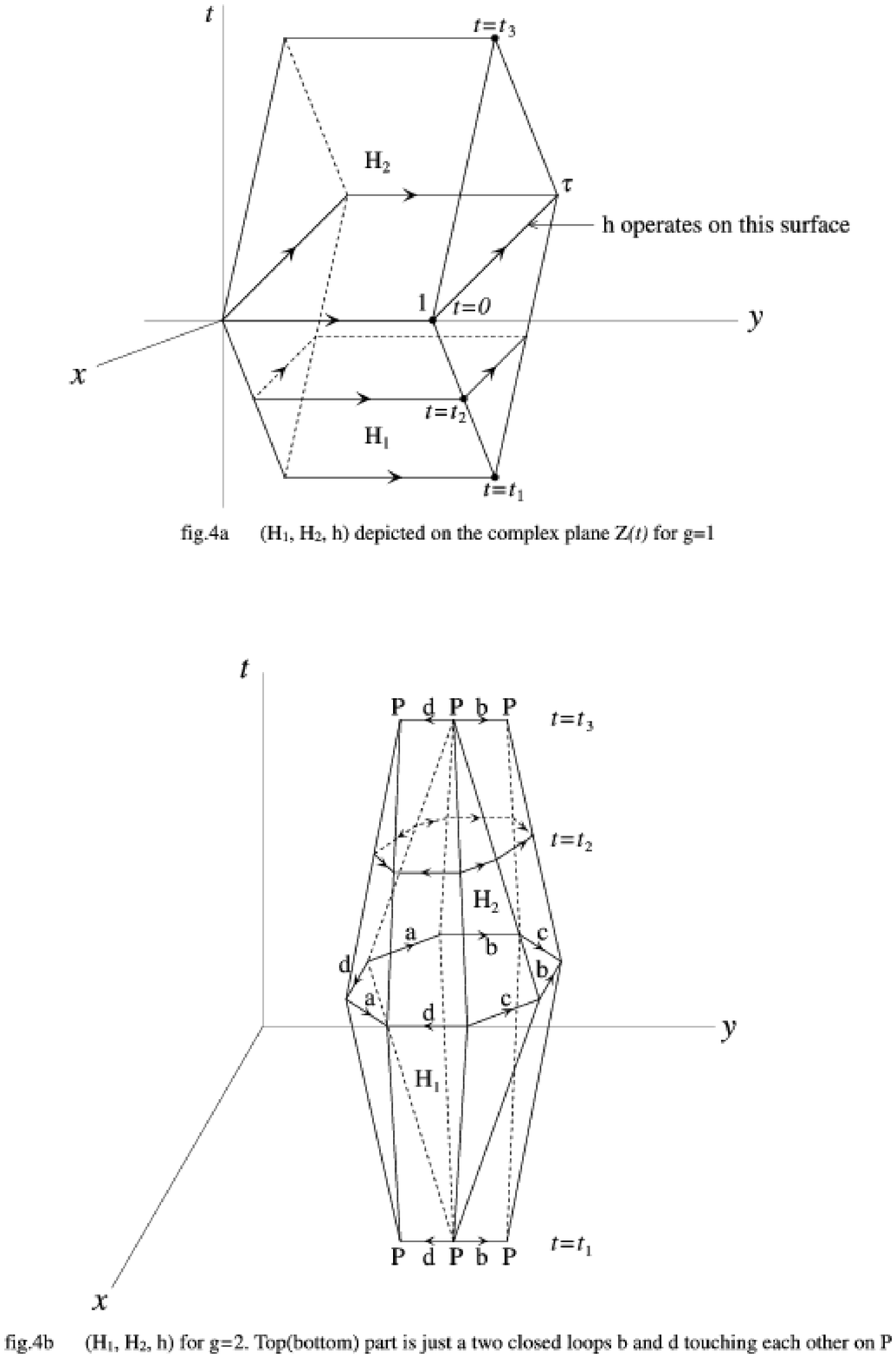}
\end{center}
\end{figure}

\begin{figure}[htbp]
\begin{center}
\includegraphics[width=16cm]{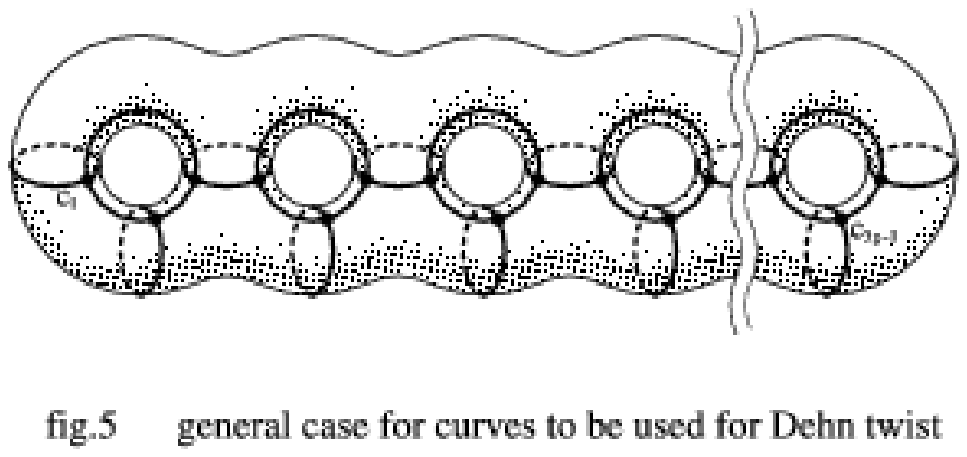}
\end{center}
\end{figure}

\begin{figure}[htbp]
\begin{center}
\includegraphics[width=16cm]{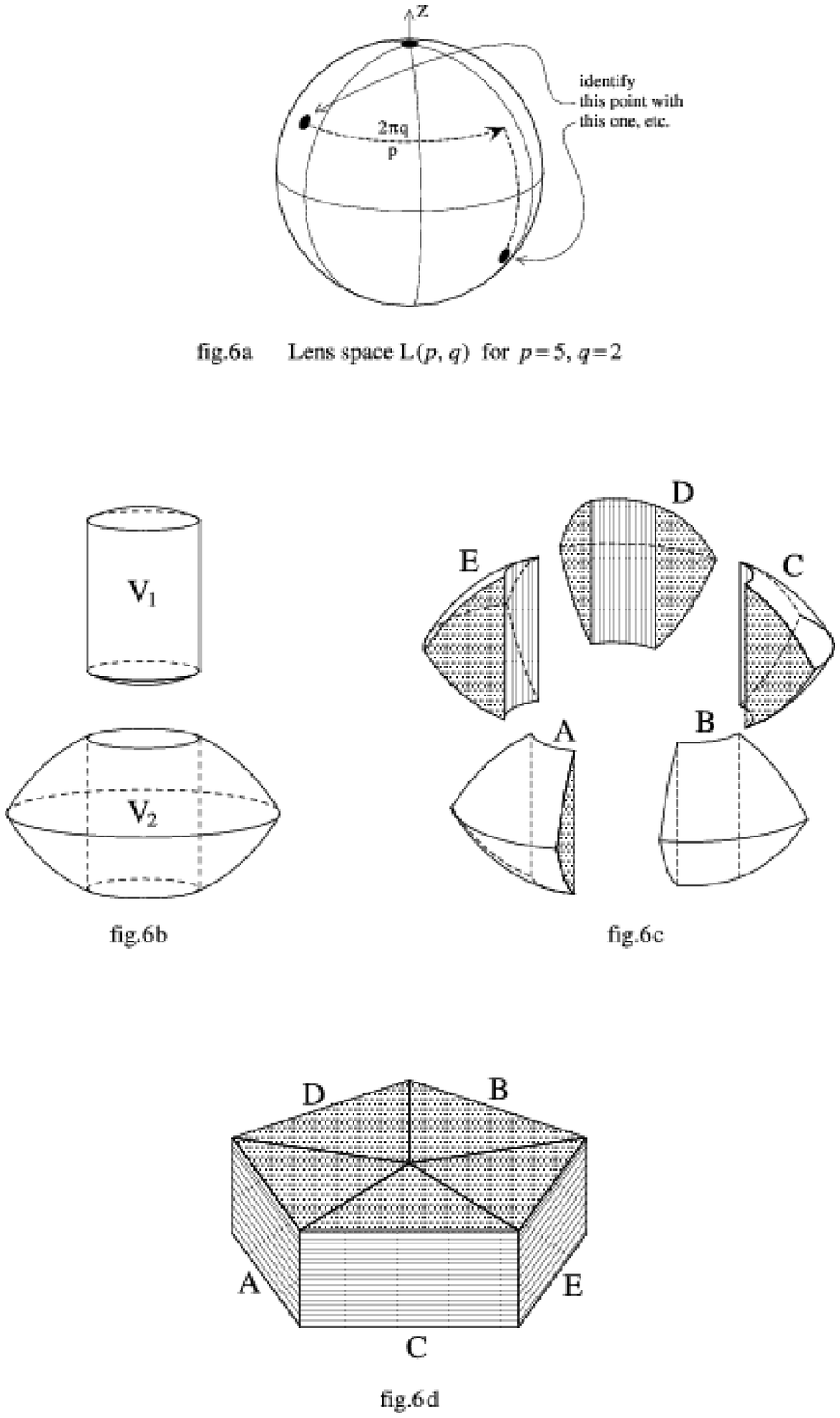}
\end{center}
\end{figure}

\begin{figure}[htbp]
\begin{center}
\includegraphics[width=16cm]{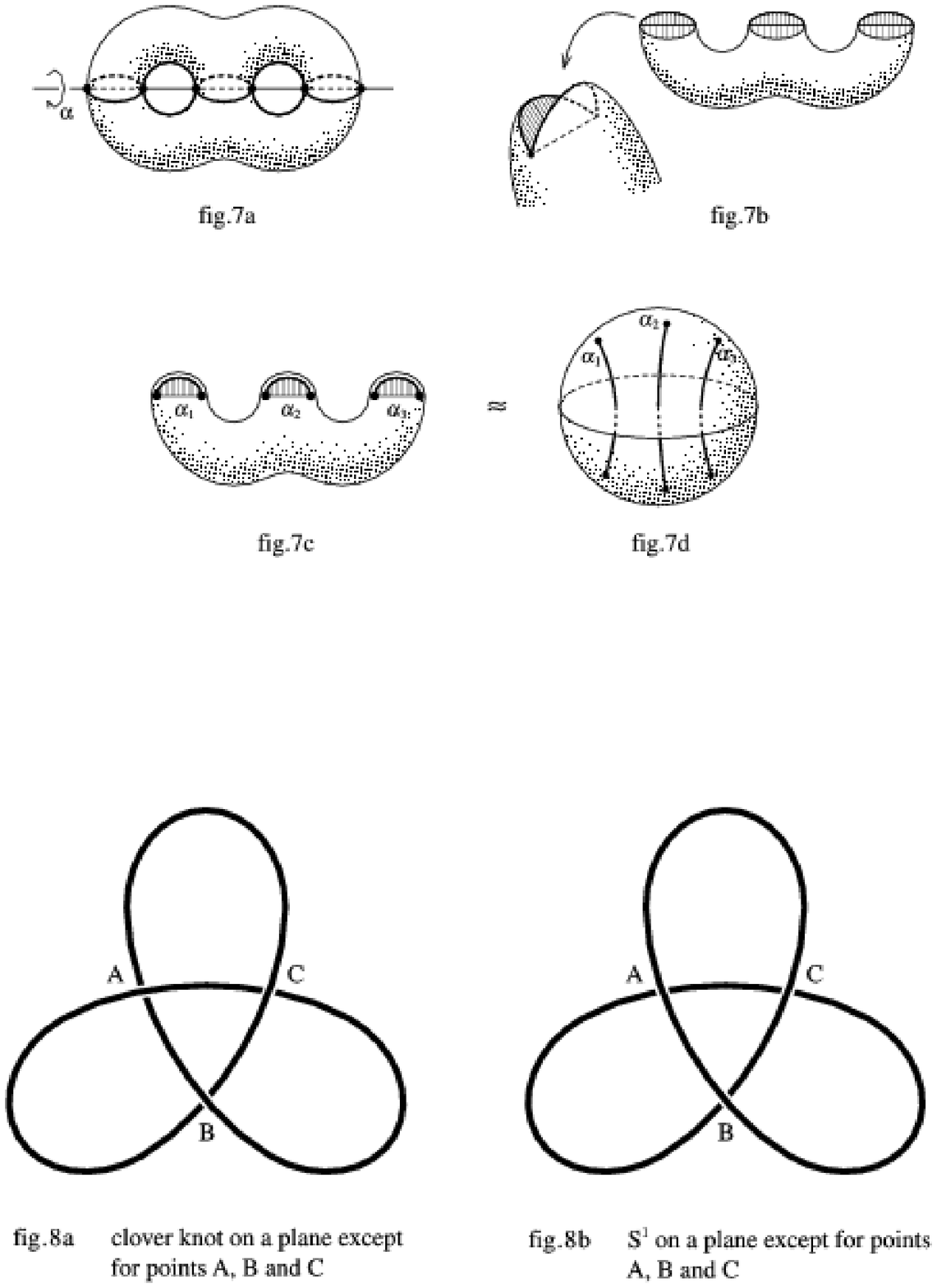}
\end{center}
\end{figure}

\end{document}